\date{}
\documentclass[showpacs,floatfix]{revtex4}
\usepackage{graphicx,psfrag,amsmath,amssymb,amsfonts,bbm,latexsym,color,dcolumn,epsf}

\begin{document}
\title{Visibility Fringe Reduction Due to Noise-Induced Effects:
Microscopic Approach to Interference Experiments}

\author{Paula I. Villar \footnote{paula@df.uba.ar}}
\author{Fernando C. Lombardo\footnote{lombardo@df.uba.ar}}
\affiliation{Departamento de F\'\i sica {\it Juan Jos\'e
Giambiagi}, Facultad de Ciencias Exactas y Naturales, UBA; Ciudad
Universitaria, Pabell\' on I, 1428 Buenos Aires, Argentina}

\date{today}

\begin{abstract}
Decoherence is the main process behind the quantum
to classical transition. It is a purely quantum
mechanical effect by which the system looses its ability to exhibit
coherent behavior. The recent experimental observation of
diffraction and interference patterns for large molecules raises
some interesting questions. In this context, we identify possible
agents of decoherence to take into account when modeling these
experiments and study theirs visible (or not) effects  on
the interference pattern. Thereby, we present an analysis of matter
 wave interferometry in the presence of a dynamic quantum
environment and study how much the visibility fringe
 is reduced and in which timescale the decoherence effects
destroy the interference of massive objects. Finally, we apply our
results to the experimental data reported on fullerenes and cold neutrons.

\end{abstract}

\maketitle


\newcommand{\beq}{\begin{equation}}
\newcommand{\eeq}{\end{equation}}
\newcommand{\dalam}{\nabla^2-\partial_t^2}
\newcommand{\mbf}{\mathbf}
\newcommand{\itm}{\mathit}
\newcommand{\beqa}{\begin{eqnarray}}
\newcommand{\eeqa}{\end{eqnarray}}

\section{Matter Wave Interferometry and Decoherence}
Matter wave interferometers are based on quantum superpositions
 of spatially separated states of a single particle. However, as
 is well known, the concept of wave-particle duality is not
 applicable to a classical object because this kind of object
 never occupies macroscopically distinct states simultaneously.
 Then, by performing interference experiments with massive particles,
 in particular with those of increasing complexity, one can probe
 the borderline between these incompatible descriptions and
shed some light on one of the corner stones of quantum physics.

Matter wave interferometry has been largely studied in the last
few years. Many theoretical studies have been done around the
mesoscopic systems \cite{Facchi,Brezger}. Mesoscopic objects are
neither microscopic nor macroscopic. They are generally systems
that can be described by a wavefunction, yet they are made up of a
significant number of elementary constituents, such as atoms.
Well-known examples these days are fullerene molecules $C_{60}$
and $C_{70}$, which are expected to behave like classical
particles. Nonetheless, the quantum interference of these
molecules has been observed \cite{Horn2}. In these experiments,
also done with cold neutrons, thermally produced beams are
collimated, diffracted by a grating, and then detected on a
distant screen. The pattern so produced shows a
typical interference profile of wave phenomena slightly
attenuated.

Usually, the main problem in the analysis of interference 
experiments is to establish exactly which are the causes for the
loss of spatial coherence observed in the reduction of the visibility
fringe of the interference pattern therein.  Macroscopic quantum
states are never isolated from their environments \cite{Zurek}.
They are not closed quantum systems, and therefore, they cannot
behave according to the unitary quantum-mechanical rules.
Consequently,
 these so often called ``classical"
systems suffer a loss of quantum coherence that is absorbed by the
environment. This decoherence destroys quantum interferences. For
our everyday world, the timescale at which the quantum
interferences are destroyed is so small that, in the end, the
observer is able to perceive only one outcome, i.e. a classical
world. As far as we see, decoherence is the main process behind
the quantum to classical transition. Formally, it is the dynamic
suppression of the interference terms induced on subsystems due to
the interaction with an environment.

In principle, some incoherence (lack of coherence) 
effects  can be imputed to the passing of the particle through the slits, such as 
vibrations or Van der Waals interactions \cite{Grisenti99} or the
difference in size of the slits \cite{Bozic04}. In the present work 
we shall not consider such effects as they are in general negligible under 
suitable experimental conditions. We shall  
consider experiments where coherent states of massive particles 
are well prepared and the diameter of the particles are smaller 
than the width of the slits in order to avoid the consideration 
of the above mentioned 
effects.

In matter wave interferometry experiments, several losses of
spatial coherence affect the particle beam during its evolution, 
with a consequence of a 
fringe visibility reduction
of the detected intensity pattern. These dynamic decoherence 
effects can be
imputed to collisions with the air
molecules or thermal photons, for example. Formally, 
decoherence appears as
soon as the partial waves (the wavefunction of the subsystem, i.e.
massive particle) shift the environment into states orthogonal to
each other. However, the loss of spatial quantum coherence can
alternatively be explained by the effect of the environment over
the partial waves, rather than how the waves affect the
environment. It is a consequence of the entanglement between the system 
and its environment. The loss of spatial coherence can also be 
originated in the angular divergence,  the
non-monochromaticity of the beam and the randomness in the
emission or arrival of the particle (mainly related to the
experimental difficulty in the production of the same initial
state for all the particles). This randomness gives raise to a
fluctuating phase $\phi$ and therefore, the interference term
appears multiplied by a factor $e^{i\phi}$. The effect can be
directly related to the statistical character of $\phi$, in
particular in situations where an external potential exerted
on the partial waves is not static. We associate these effects to the {\it
dephasing} process. Yet
more importantly, any source of stochastic noise would create a
decaying coefficient. In this way, the uncertainty in the phase
produces a decaying term that tends to eliminate the interference
pattern. This quantum suppression is due to the presence of a noisy
environment coupled to the system and can be represented by
 the Feynman-Vernon influence functional formalism \cite{Stern}.

Nonetheless, it is relevant to explain the
quantum-to-classical transition in a unified framework since the
understanding of the decoherence (or dephasing) phenomena points out
the crucial role played by the environmental interaction in
determining whether a quantum particle shows wave behavior. Thus,
there is a need to theoretically quantify the effect of decoherence
(or dephasing)  on the observed interference pattern.  It is quite 
intuitive that the resulting
pattern shall be an interplay between the strength of the coupling
to the environment, the slit separation and the distance the
particle travels from the slit to the screen. The decoherence
effects on two-slit experiments have been theoretically analyzed by
treating the effect of the environment using a phenomelogical model
in \cite{Sanz,Facchi}. In \cite{Venugopalan}, authors described
theoretically the effects on the interference pattern assuming the
test particle develops a quantum brownian motion and solving the
corresponding master equation, neglecting in the end the dissipation
of the environment on the system. Contrary to these studies, authors
in \cite{tumulka} stated that the dynamic decoherence does not play any role in
the visibility fringe reduction and blamed the latter on the
incoherence of the source.

In the present paper we study the visibility fringe reduction in the
interference pattern of experiments with
particles, such as fullerenes and cold neutrons. The questions to be
addressed are:  how long can we observe before decoherence or
dephasing effects destroy the interference pattern of massive
particles? How much the visibility fringe is reduced in these
experiments and which are the possible agents of decoherence to take
into account when modeling these experiments? Therefore, in this
paper we shall study both the dephasing effects due
to a random variable (in our case the particle's emission time) and
the dynamic decoherence process obtained from a first principles
model. Even though phenomenological models of environmental decoherence 
success fitting experimental data, we stress that a complete 
description of the interaction between system and environment is needed 
in order to get a well defined quantum to classical transition.

The paper is organized as follows. In Section \ref{teoria} we
present the different decoherent agents that can be used to model
 this type of experiments and develop the theoretical frames to
study how these agents affect the interference pattern. In Section
\ref{numerico}, we introduce the numerical tools used in order to
quantify the visibility fringe reduction in the pattern
of an interference experiment. This is done using both
analytical and numerical results. Section \ref{aplicacion} contains
an application of the models described in the previous sections to
real matter wave interferometry experiments performed with cold
neutrons. Finally, in Section \ref{final}, we include our final
remarks.

\section{Theoretical Analysis}
\label{teoria}

\subsection{Two Gaussian localized wave packets}

We shall study a typical interference experiment with
particles of mass $M$ diffracted by a grating and then detected on
a distant screen. The particle leaves
the grating and travels a distance $L$ in the y-direction until it
reaches the screen in a time $t_L=M L/p_0$, where $p_0$ is the
moment's component in that direction. It is important to note that,
in order to observe an interference pattern on the screen, particles
should be coherent in the x-direction, whereas, the dynamics in the
y-direction can be that of a free non-interacting particle. Hence,
the experiment starts by the preparation of the initial state that
emerges from the slits. Initially, we may reasonably assume
that we have a coherent superposition of the
two wave packets, centered at each location of the respective slits
and factorized as \cite{Venugopalan,Viale}\[
\Psi(\vec{x},0)= (\phi_1(x,0) + \phi_2(x,0)) \otimes \chi(y,0),
\] where $|\phi_1|^2$ and $|\phi_2|^2$ correspond to the probability
amplitudes for the particle to pass through slit $1$ and slit $2$
(in the x-axis), respectively, while $\chi(y,t)$ represents the
Gaussian wave function in the y-direction (where no superposition is
needed). Note that we are assuming translational invariance in the
z-axis \cite{Viale}.

The interference pattern, in any case, corresponds to the
probability distribution of the time evolved wave function: \beq
P(\vec{x},t)= \bigg( \phi_1(x,t)^* \phi_1(x,t)
+\phi_2(x,t)^* \phi_2(x,t) +\phi_2(x,t)^*
\phi_1(x,t) 
+\phi_1(x,t)^* \phi_2(x,t) \bigg)|\chi(y,t)|^2,  \eeq which is 
 the diagonal part of the density matrix defined as
$\rho(\vec{x},\vec{x}',t)= |\Psi(\vec{x},t)\rangle \langle
\Psi(\vec{x}',t)|$.

When the system is closed, the quantum states of the system evolve
accordingly to the Schr\"odinger equation. In such a case, it is easy
 to show that the position
probability distribution on the screen at a given time $t$ is:
\beq
P(\vec{x},t)= \bigg(|\phi_1(x,t)|^2+ |\phi_2(x,t)|^2 
+ 2 {\rm Re} (\phi_1^*(x,t) \phi_2(x,t)) \bigg)
 |\chi(y,t)|^2. \nonumber
\eeq However, when the system is open, it interacts with an
environment and its evolution is plagued by nonunitary features like
{\it fluctuations} and {\it dissipation}, no matter how weak the
coupling that prevents the system from being isolated is.
Particularly, {\it decoherence}, as we mentioned in the preceding
section, is the dynamic suppression of the interference terms
induced on subsystems due to the interaction with an environment.
For a superposition of localized wavepackets (which best describe
massive particles), the initial ($t=0$) four terms of the density
matrix $\rho(x,x',0)=\phi_1(x,0)^* \phi_1(x',0)+ \phi_2(x,0)^*
\phi_2(x',0)+\phi_2(x,0)^* \phi_1(x',0)+\phi_1(x,0)^* \phi_2(x',0)$
correspond to four peaks. Decoherence arguments show that the
off-diagonal terms die out due to the interaction with the
environment. As the interference pattern depends on the {\it
diagonal} components of the density matrix, it is not obvious if the
suppression of the coherences of the density matrix due to the
decoherence process also corresponds to a disappearance of the
interference pattern. In the case of open systems, the object of
study is the reduced density matrix $\rho_r(x,x',t)$ of the
subsystem (massive particle) which satisfies a master equation (see
below).

Initially, we can assume that the environment and the total wave
function of the system factorizes as $\Psi(\vec{x},0)= [\phi_1(x,0)
+\phi_2(x,0)]\chi(y,0)
\zeta(\vec{X},0)$, 
where we have introduced a new wave function  $\zeta(\vec{X},t)$
 to describe the
state of the environment. The interference pattern at a given time
$t$ on the screen is now given by:\beq P(\vec{x},t)= \rho_r(x,x,t)|\chi(y,t)|^2
=\bigg(|\phi_1(x,t)|^2+ |\phi_2(x,t)|^2 
+ 2 {\bf \Gamma}(t){\rm Re} (\phi_1^*(x,t) \phi_2(x,t))\bigg)
|\chi(y,t)|^2
\label{gamma} \eeq where $\Gamma(t)$ encodes the information about
the statistical nature of noise since it is obtained after tracing
out the degrees of freedom of the environment. It is, in general, an exponential
decaying coefficient which suppresses the interference terms in a
decoherence time scale $t_D$. It is important to stress that in this
overlap factor $\Gamma(t)$ we can include not only the dynamical decoherence
effects but also the dephasing ones induced on the subsystem due
to a coupling to an external reservoir \cite{Stern}.

\subsection{Different decoherent agents}
\label{intro} 

In order to complete the analysis, we need to identify
the possible ``decoherent" agents so as to estimate the overlap
factor $\Gamma(t)$ for the different types of environment considered
when modeling a two slit experiment. In the literature
there can be found many studies that blamed the reduction of the
visibility fringe on different causes: from the irregularities of
the grating (these are named incoherence effects, and they are not really dynamical 
decoherence since they are related with the source or the 
preparation of the initial state) 
to the scattering of the massive particles with the air
molecules to the dephasing generated by the collimation of the
beams.

A valid assumption, although a rather simplified version of the real
problem, is the implementation of the model of Joos and Zeh,
hereafter called scattering model \cite{Joos}, in order to study the
dynamics of the test particles moving in a quantum medium. This
model, which basically is a phenomenological description of
processes inducing loss of coherence in a quantum system,
considerers that the reduced density matrix of the system evolves
autonomously according to a markovian-type master equation (see also Refs.
\cite{Walls85,Diosi95})
\beq i
\frac{\partial \rho_r}{\partial t} = [H,\rho_r] - i \Lambda
[x,[x,\rho_r]].\label{scatering} \eeq The effect of the environment
is summarized by a collision term, added to the free dynamics of the
system, which takes into account the {\it decoherence} in the
coefficient $\Lambda$ but neglects {\it dissipation} (see discussion in 
\cite{Vacchini00,Vacchini01}). As an example,
in Ref.\cite{Viale}, authors consider $\Lambda= \Lambda_{\rm{air}} +
\Lambda_{\rm{photons}}$, and state that the cause of decoherence in
this type of experiments might be the scattering of the particles
with air molecules and thermal photons during their flight from the
slits to the screen \cite{Gallis90}. Eq.(\ref{scatering}) corresponds to the 
many scatterer or high temperature limit of a more general equation \cite{Pritchard}. Consequently, 
for this model, the effect of the
environment is encoded in $\Gamma(t)=\exp(-\Lambda t)$ ($\Lambda$ is
phenomenologically estimated through the wavelength and scattering
cross section of the particles) and only considers the dynamic
monitoring of the environment over the subsystem (i.e. dynamic
decoherence base on a phenomenological model) \cite{Pritchard}. As 
the contrast of the interference pattern is proportional to the coherence 
between the two paths, reduction in the contrast will be a direct indicator  
of decoherence. Therefore, spatial coherence loss of a superposition state is 
due to scattering events. In the many scatterer limit, Eq.(\ref{scatering}) agrees 
with data. Thus, decoherence is exponential with time and with the path 
separation squared, as decoherence theory usually predicts.
In this context, other decoherence models can be applied, as the one 
by Hornberger, Sipe, and Arndt \cite{Horn} which uses of the phase 
space description provided by the Wigner function to explain 
decoherence effects in a matter wave Talbot-Lau interferometer; or 
the more recent works by Hornberger on the formulation of the master 
equation for a quantum particle in a gas \cite{Horn2}. Even though we shall not 
considerer the Fraunhoufer limit, it is important 
to note that in Ref.\cite{Horn3} thermal limitation of far-field interference 
has been reported. 

Another approach, which is a dephasing model, might be to consider
the influence of the external classical time-dependent
electromagnetic field on the experiment as we have previously done
in \cite{Nos}. The interaction between the particles (electrons or
neutral particles with permanent dipole moments) and classical
time-dependent fields induces a time-varying Aharonov phase.
Therein, we included a random variable $t_0$, which is defined as
the particle's emission time. This variable produces a fluctuating
phase $\phi$ which averaged in time produces a decaying term that
reduces the fringe visibility of the interference pattern. In this
way, the uncertainty in the phase originates decoherence effects
caused by the experimental difficulty of producing the same emission
time for all particles and estimated as \beq
F = \langle e^{i\phi}\rangle 
= \lim_{T\rightarrow \infty}\frac{1}{2T}
\int_{-T}^{T}dt_0 \exp\{i[A \cos (\omega t_0)  
+ B \sin (\omega t_0) ]\} = J_0(\vert C\vert ),\nonumber \eeq
where $J_0$ is the Bessel function. The modulus of complex number
$C = A + i B$ measures the degree of dephasing. The overlap factor
F encodes the information about the statistical nature of noise.
Therefore, classical or quantum noise makes F less than 1, and the
idea is to quantify how slightly it destroys the particle
interference pattern.
 Hence, in this case, $\Gamma \equiv F$. Notably
in this approach, the effect of the
environment is constant through all the experiment since $\Gamma$
is obtained after averaging in time and therefore, does not depend
upon time.

Finally, as we previously said, generally, the passage of the
particles through the grating can produce vibrations, or other kind
of interactions with the walls of the grating, able to corrupt the
visibility of the interference pattern due to alterations in the 
initial coherence of the superposition. Moreover, also the finite
size of the grating and the differences in the slit aperture can
attenuate the visibility of the interference fringes, especially in
the case of complex (very large) molecules
 \cite{Horn}. In the present article, we shall only concentrate on modeling 
the interaction of the interfering 
particles and their environment, from a microscopic quantum level. In this case, 
the dynamics of the test particles
 can be modeled
by the quantum brownian motion (QBM)  \cite{Hu} and the reduced
density matrix of the system satisfies a master equation (see
Eq.(\ref{master}) below) with the diffusion coefficient ${\cal D}(t)
= 2 M \gamma_0 k_B T$ for ohmic environment in the high temperature
limit (when experiments at room temperature are made with large
molecules, i.e. fullerenes, and cold neutrons this last
approximation is valid). Not only is the diffusion considered in
this model environment but also the dissipation (through the
coefficient $\gamma(t)$). Then, in this case, $\Gamma(t)=\exp(-{\cal
D}t)$ represents the noise induced environmental effect on the
system due to the interaction with the environment. Scattering models 
or no-damped motion are just approximations obtained from our 
general framework \cite{Ambegaokar}. More especulative type of environments can 
be considered, such as 
space-time foams, quantum gravity effects, etc; but they are out the scope of 
our work since there is no experimental evidence of such decoherence 
agents on matter waves (see for example \cite{Nik,Lamine}).

\section{Numerical Analysis}
\label{numerico}

\subsection{Interference pattern}

In this Section, we shall study the interference pattern produced by
two well localized Gaussian wave packets, initially given by, \beq
\Psi(\vec{x},0)= N \bigg( \exp(\frac{(x-L_0)^2}{4 \sigma_{x0}^2})
+ \exp(\frac{(x+L_0)^2}{4 \sigma_{x0}^2}) \bigg)
   \exp(-\frac{y^2} {4 \sigma_{y0}^2}-i k_y y) \label{2gaus}
\eeq where $2L_0$ is the initial separation of the center of the wave packets,
$\sigma_{x0}^2$  and
 $\sigma_{y0}^2$ are the initial
width of the packet in the x and y-axis, respectively, and $k_y$ the
initial moment of the particle in the y-direction. It is important
to note that $L_0$, $\sigma_{x0}$, $\sigma_{y0}$ and $k_y$ are all
free parameters that have to be tuned with the experimental data. In
addition, we assume that $\Delta p_y << p_y$, so the moment
component is sharply defined and the wave packet has a
characteristic wavelength $\lambda_{dB}$ associated $\lambda_{dB}
\sim \hbar/p_y << \Delta y$.

We shall study the effect of decoherence on the interference pattern
of an experiment with massive particles, by
coupling our subsystem (particles) to a model environment. As we
mentioned above, the experiment consists of massive particles
(represented by the superposition of two localized wave packets)
that are diffracted by a grating and registered later on a screen at
a distance $L$. As we have already stated, the dynamics in the
y-direction just serves to transport the particles from the slit to
the screen and can then be considered as a ``free" evolution. However,
in the x-direction we need to consider a decoherent agent in order
to study the effect of decoherence on the interference pattern
observed on the screen. Thus, hereafter, we shall consider that the
environment is a set of non-interacting harmonic oscillators and the
dynamics of the test particles is modeled by a QBM. As noted in the 
preceding section, this behavior can be
blamed on any interaction by which the particles-system become 
entangled with a quantum environment. In order to study the interference pattern registered
on the screen at a later time $t_L$, we need to obtain the evolution
in time of the reduced density matrix $\rho_r(x,x',t)$, which is
given by the following master equation \beq \frac{\partial
\rho_r}{\partial t} = \frac{i \hbar}{2 M} \bigg(
\frac{\partial^2 \rho_r}{\partial x^2}- \frac{\partial^2 \rho_r}
{\partial x'^2}\bigg) 
- \frac{{\cal D}(t)}{4 \hbar^2} (x-x')^2 \rho_r 
- \gamma(t) (x-x') \bigg(\frac{\partial \rho_r}{\partial x}-
\frac{\partial \rho_r}{\partial x'} \bigg) 
+ 2 f(t) (x-x') \bigg(\frac{\partial \rho_r} {\partial x} +
\frac{\partial \rho_r}{\partial x'}\bigg), \label{master} \eeq
where  $\gamma(t)$ is the dissipative coefficient (proportional to
the square of the coupling constant to the environment), ${\cal
D}(t)$ the diffusive coefficient and $f(t)$ the coefficient
responsible for the anomalous diffusion. Eq.(\ref{master}) has been
obtained by assuming the environment to be in equilibrium, at a
temperature T.  In the case that the system is coupled to an ohmic
environment in the high temperature limit ($k_B T >> \hbar \omega$),
these coefficients are constant $\gamma(t)=\gamma_0$, ${\cal D}(t) =
2 M \gamma_0 k_B T$ and $f(t) \approx 1/k_B T$ \cite{Hu}. We restrict 
ourselves to the use of the ohmic bath since it is the type of 
environment which produces the correct limit for classical 
dissipation. It is the most studied case in the literature and produces 
a dissipative force that in the limit of the frequency cutoff 
$\omega_{\rm cutoff}\rightarrow 0$ 
is proportional to the velocity. In order to model more complex 
interactions (like charges with fields) it could be more appropriate 
to use a supraohmic spectral density. Nevertheless, it is well known 
that dynamic decoherence in the high temperature limit ocurs 
in a similar time-scale both for ohmic and supraohmic environments 
\cite{Hu,Phases}. This is 
the reason why we shall only concentrate on the simplest case. General 
type of environmnets can be easily included in our approach, but it is not 
possible to get Eq.(\ref{scatering}) as a limit from (\ref{master}) for 
general non-ohmic environments.  

It is important to stress that Eq.(\ref{scatering}) can be obtained
from Eq.(\ref{master}) in the high temperature limit {\it of an ohmic environment} 
(neglecting
dissipation) for the markovian case if written in the Lindblad form.
However, master equation Eq.(\ref{master}) refers to a more general movement that
can be used for all temperatures and spectral densities, even to study the dynamics of the
test particle at zero temperature (non-Markovian) limit \cite{PLA,PRE}. Yet more interesting, this
formulation  verifies the
fluctuation-dissipation theorem for a general system in thermal
equilibrium \cite{Ambegaokar}. It is also important to note that the high temperature
limit approximation is well defined only after a time scale of the
order of $1/(k_BT)\sim \gamma_0/{\cal D}$ ``ensuring" the 
positivity of the reduced density matrix $\rho_r(x,x',t)$ \cite{Vacchini00}.

Thus, as we mentioned above, in order to study the dynamics of these
two packets that best describes the massive particle, we need to
solve the master equation Eq.(\ref{master}). The corresponding
density matrix that arises from the initial state given by
Eq.(\ref{2gaus}) is \beq \rho_r(\vec{x},\vec{x}',0)=
\rho_r(x,x',0) \otimes  \rho_r(y,y',0)
=  2 N^2 \bigg(\cosh(2 L_0 (x+x')) + 
 \cosh(2 L_0 (x-x')) \bigg)\chi(y,0)^*\chi(y',0). \nonumber \eeq
The solution in the x-direction can be well reproduced by employing
a Gaussian density matrix using the Born approximation \cite{Viale,Joos}. Is is 
worth noting that this solution does not imply far-field or Fraunhofer 
approximation.
\beq
\rho_{\rm r}(x,x',t)= e^{-N(t)} \exp \bigg\{
 - A(t) (x-x')^2
-  i B(t)(x^2-x'^2) - C(t) (x+x')^2  \bigg\} \label{1gaus}
\eeq
where $e^{-N(t)}$ ensures the conservation of trace, $A(t)$
describes the range of coherence while $C(t)$ specifies the
extension of the ensemble in space. All functions $A(t),~B(t),$...
are real for the sake of hermicity. In our case, we shall
study the dynamical evolution of two Gaussian wave packets located
at $x=\pm L_0$. Therefore, we have to replace $x\rightarrow x+L_0$
and $x\rightarrow x-L_0$ in Eq.(\ref{1gaus}) and superpose both
anzats in order to represent the dynamics of the two packets. In
that case, the solution we shall use is
\beqa
\rho_{\rm r}(x,x',t)&=& 2 e^{-N(t)} e^{-4 L_0^2 C(t)}
\exp \bigg\{ - A(t) (x-x')^2  
- i B(t)(x^2-x'^2)- C(t) (x+x')^2 \bigg\} \times \nonumber \\
&&  \bigg(  \cosh [4 L_0 C(t) (x+x') 
- i 2 L_0 B(t) (x-x') ] \nonumber \\
&+& e^{-4 L_0^2 (A(t)-C(t))} \cosh [4 L_0 A(t) (x-x')
+ i 2 L_0 B(t) (x+x') ] \bigg).
\label{2gausrho}
\eeqa

We have numerically solved Eq.(\ref{master})  for a free particle
assuming its dynamics is modeled by QBM (in the x-axis) using a
standard adaptative-step-size fifth-order Runge-Kutta method with
initial condition $A(0)=1,~B(0)=0,~C(0)=1$ in units of $\hbar=1=M$.
By doing this, we obtained the dynamic evolution of the coefficients
$A(t),~B(t)$, $C(t)$ and $N(t)$. All results were found to be robust
under changes in the parameters of the integration method.

The intensity registered on the screen at a given time $t$ is
proportional to the position probability (diagonal term of the
reduced density matrix) $ P(x,t) \approx \rho_r(x,x,t)$. In the case
of the initial state mentioned above, the intensity can be
numerically obtained \beq P(x,t)= e^{-\tilde{N}(t)} e^{-4C(t)
(x^2-L_0^2)}\bigg(\cosh(8 C(t)
L_0 x) 
+ \Gamma(t) \cos(4 B(t) L_0 x) \bigg), \label{prob} \eeq where we
have absorbed the decaying term coming from the Gaussian wave in the
y-direction $|\chi(y,t)|^2$ in the normalization
$e^{-\tilde{N}(t)}$, and $\Gamma(t)$ is a decaying exponential $
\Gamma(t)=e^{-4 L_0^2 (A(t)-C(t))}$ (provided $A(t)-C(t) >0$). We know  
the dynamic evolution of the interference
pattern at a  distance $L$ in
the case the system is isolated. The two initial
wave packets start to evolve in time and spread in the x-direction.
Immediately, they start to develop an interference pattern. In the
case of the system is interacting with a very strong environment, 
it is clear that for the same times (or even shorter ones)
the interferences can not be observed because they are almost
immediately destroyed. As the evolution continues (for a fixed
length of the screen), the two packets continue spreading. After
some time, we can no longer observe two packets on the screen
because both wave turned into one (because of the spread of each
packet). For this type of environment no interferences fringes will
be observed for these thought experimental times.

\subsection{Estimation of the Decoherence Time and
Fringe Visibility Reduction}

\begin{figure}[!h]
\includegraphics[width=17.cm]{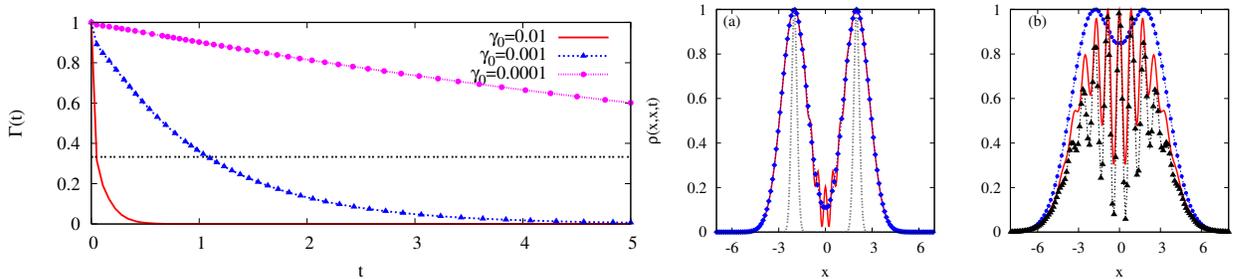}
\caption{On the left: Evolution in time for the decaying exponential 
$\Gamma(t)$ that destroys the interferences of the system for different
couplings to an environment in the high temperature limit.
We use units: $\hbar$ = c = M = 1. 
Parameters are: $L_0=2 s^{-1}$, $\sigma_{x0}=0.5 s^{-1}$,
$k_B T=300 s^{-1}$. The stronger the coupling to the environment 
$\gamma_0$ is, the sooner decoherence effects take place 
(for a fixed value of $k_B T$ and $L_0$). On the right: 
Interference pattern registered on the screen at a time
$t_L$ for the closed and open system. In (a), we have considered the
case of the isolated subsystem (solid red line) and the case of
coupling to a strong environment with $\gamma_0=0.01~s^{-1}$ (blue dot
line). In (b), we have considered three different environments:
$\gamma_0=0.01~s^{-1}$ (blue dot line),  $\gamma_0=0.001~s^{-1}$
 (red solid line) and $\gamma_0=0.0001~s^{-1}$ (black triangle line). Distance 
is measured in units of frequency.} \label{graf1}
\end{figure}

We shall estimate the decoherence time $t_{{\cal D}}$, i.e. the timescale for
which the interferences are mostly destroyed, as $\Gamma_{{\cal
D}}(t_{{\cal D}})=\exp(-{\cal D}\Delta x^2 t_{{\cal D}}) \sim 1/e$.  It is easily
deduced that $t_D \approx 1/({\cal D} \Delta x^2)$, with $\Delta
x^2=(x-x')^2$ and ${\cal D}=2 M \gamma_0 k_B T$ for the ohmic
environment in the high temperature limit in units of $\hbar=1$, as
shown on the left side of  Fig.\ref{graf1}.\\

Clearly, since the decoherence timescale depends inversely on the
value of $M \gamma_0 k_B T$, the stronger the coupling to the
environment and the hotter the environment, the shorter this
timescale.

On the right side of Fig.\ref{graf1}, we can see the effects of decoherence on the
interference pattern
 of a thought two-slit interference experiment with particles.  In plot
(a), we show the interference pattern registered on a
screen at a distance $L$ in a time $t_L=0.2~s$ when the system is
closed, i.e. there is no interaction with an environment, and when
the system is open. In this latter case, the coupling constant is
$\gamma_0=0.01$, which represents a strong environment because all
interferences have already been destroyed (whereas they are present
in the isolated case. In (b), we present a
latter time ($t_L=0.35~s$) for different coupling constants to the
environment. We can see that for  $\gamma_0=0.01~s^{-1}$, the two wave
packets
 are spreading and will end up superposing in only one final wave packet
since the environment has destroyed the interference in a  short
timescale. However, for the other two environments, with smaller
coupling constants, we can see that the interferences are still
there. Notably, the pattern remains unchanged but the visibility is
considerably reduced as $M \gamma_0 k_B T$ increases. It is
important to note that the visibility is considered attenuated when
there is a lost of contrast between a maximum and a minimum with
respect to the interference pattern when the system is isolated,
i.e. the visibility is reduced when the ``minimum" are not exactly
zero as seen in Fig.\ref{graf1}.

\begin{figure}[t]
\includegraphics[width=16cm]{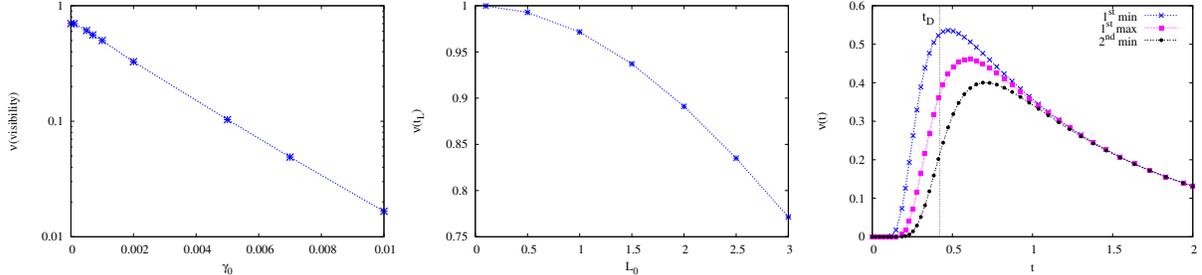}
 \caption{{\it Left}: The visibility fringe $\nu(t)$ plotted against $\gamma_0$ for
 a fixed $t_L < t_D$ at room temperature ($\gamma_0$ is in units of frequency).
{\it Middle}: The visibility fringe $\nu(t)$  as a function of the
distance between the slits $L_0$ for a fixed time for $k_BT=300~s^{-1}$,
$\gamma_0=0.001~s^{-1}$, $t_L=0.05~s$ and $\sigma_{x0}=0.5~s^{-1}$. All the
decoherence timescales corresponding to the different values of  $L_0$ (in units of $s^{-1}$)  
are checked to be longer than $t_L$. {\it Right}:
Time evolution for the  visibility fringe $\nu(t)$ for
$k_BT=300~s^{-1}$, $\gamma_0=0.001~s^{-1}$, $L_0=2~s^{-1}$ and 
$\sigma_{x0}=0.5~s^{-1}$.  The
estimation of the decoherence time $t_D \sim 1/(M \gamma_0 k_BT
L_0^2)=0.41~s$ coincides with the timescale at which the visibility
starts to decrease towards a null value.}
\label{graf2}
 \end{figure}

At this stage, it is appropriate to quantify the loss of contrast of
the interference pattern. This is done by defining a function called
fringe visibility $\nu$, a quantity of particular importance in 
matter wave interferometry \[ \nu=\frac{ I_{\rm max}-I_{\rm min}}
{I_{\rm max}+I_{\rm min}},\]
where $ I_{\rm max}$ and $I_{\rm min}$ represent the maximum and
minimum in neighboring fringes, respectively. It is easy to note
that the fringe visibility can be well approximated by \[ \nu(t) 
\sim \frac{|\rho_{\rm int}(x,x,t)|}{\rho_{11}(x,x,t)
+ \rho_{22}(x,x,t)}, \] where $\rho_{ii}=|\phi_i(x,t)|^2$, with
$i=1,2$ and $\rho_{\rm{int}}$ the interference terms. The values of
this function range between $0$ (no interference fringes) and $1$
(total visibility of the interference fringes). In our case, the
visibility fringe can be numerically obtained as
\[ \nu(t) \approx \frac{\Gamma(t)}{\cosh(8 L_0 C(t) x)}.
\]
Clearly, the visibility fringe goes down as $t_L$, i.e. the
observation time, is larger than the decoherence time $t_D$.
However, if we succeed in performing our two slit experiment in a
time $t_L < t_D$ at a fixed room temperature $k_BT$, we can see that
the visibility fringes depends on $\gamma_0$ as shown on the left picture
in Fig.\ref{graf2}.
This is so, because the decoherence time
depends inversely on the coupling constant. Not only can we check the
dependence upon the coupling constant but on the separation of the
slits as well.

In the middle of Fig.\ref{graf2}, we show
the visibility fringe as a function of $L_0$. Therein, it is clear
that the visibility fringe goes down as the distance between the
slits increases. Note that as we are plotting $\nu(t)$ for a fixed
value of $t_L$ and $\sigma_{x0}$, then we can not vary much $L_0$,
since we are always assuming that $\sigma_{x0}\leq L_0$.
We can also study the time evolution
 of the visibility $\nu(t)$, which is shown on the right side of Fig.\ref{graf2}.
Therein, we have plotted the evolution in time for the visibility of
the first and second minimum and the first maximum of the
interference pattern. The behavior exhibited is quite appealing. For
short times, the visibility increases from zero to a maximum value
because the interferences start to develop at that short timescale
but are not present at $t=0$ (since the wave packets are initially
separated and have to spread so as to generate the interferences. This 
maximum value coincides with the estimated decoherence time $t_D$. Then, the
visibility starts to decrease, since the destruction of the
interferences is taking place. Clearly, the decoherence is a dynamic
process (the continuous monitoring of the environment over the test
particles) and the estimated decoherence time is when the
interferences have been reduced about a $70~\%$, i.e.
$\Gamma(t_D)\sim 1/e$ (see Fig.\ref{graf1}).  However, that does not
mean that the wigner function will be positive by that time. If one
estimates the decoherence time as the one in which the interferences
disappear completely, the estimated timescale will be longer and one
might naively impute the loss of visibility on another cause but
decoherence \cite{tumulka}. Note that the visibility is a quantity
that measures the loss of contrast of the interference fringes.
Then, it is expected that those with the bigger contrast suffer from
this attenuation the more, as seen in Fig.\ref{graf2}. Clearly, the
observation time $t_L$ must be shorter than the decoherence time in
order to observe the interference pattern. The visibility function
$\nu(t)$, in this case, tends to zero for longer times.

It shall be interesting to study the visibility function for the
other environmental models. In the case of the scattering model, the
behavior of $\nu(t)$ as a function of the diffusion term $\Lambda$ and
the square of the width of the slit $L_0^2$ is qualitatively
similar to that of the QBM because the expression of
$\Gamma_{\Lambda}(t)=\exp(-\Lambda \Delta x^2 t)$ is formally the
same. Then, we expect to find that the visibility decreases as
$\Lambda$ and $L_0^2$ increases, since the decoherence time shall
be shorter \cite{Pritchard}.

\begin{figure}[t]
\includegraphics[width=15cm]{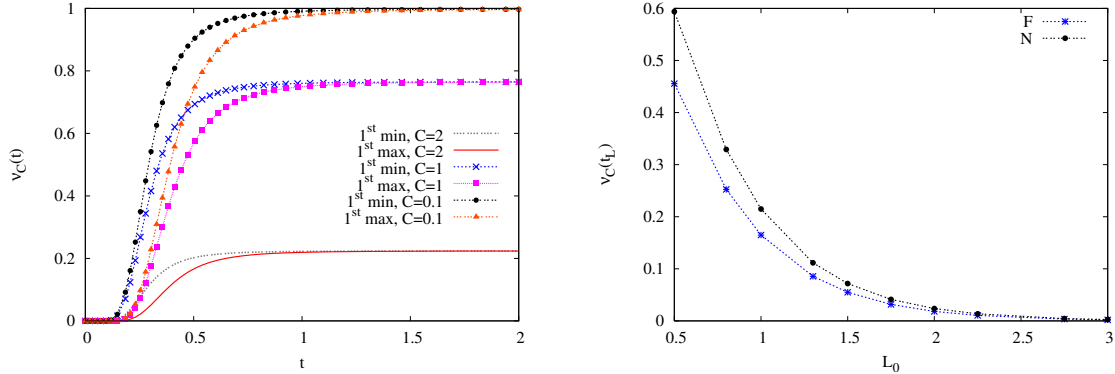}
\caption{{\it Left}: Time evolution for the visibility function $\nu_C(t)$ for
neutrons ($C_{\rm neutrons}=0.1$) and fullerenes ($C_{\rm
fullerenes}=1$ and $C_{\rm
fullerenes}=2$) in the presence of an external time dependent
electromagnetic field. The curves are for the first minimum and
maximum of the interference pattern. We can see that all curves
reach an asymptotic limit that is not zero contrary to the other
two environmental models (see Fig.\ref{graf2}). {\it Right}:
The visibility fringe $\nu_C(t)$  as a function of the
separation of the slits $L_0$ for a fixed time for time $t_L=0.20~s$,
$t_L=0.05~s$ and $\sigma_{x0}=0.5~s^{-1}$. The curves shown are for the
first maximum of the interference pattern in the case of neutrons
and fullerenes. $L_0$ is in units of $s^{-1}$.} \label{graf3}
\end{figure}

Nonetheless, the visibility function for the study of the
{\it dephasing} effects, i.e. when considering the
interaction of the massive particle with the external time dependent
electromagnetic field, is not that similar to the other two
mentioned throughout the paper. In particular, $\Gamma_C=J_0(|C|)$
is constant in time as we estimated it in \cite{Nos} for the
experimental data of both neutrons and fullerenes. Therein, we
calculated
 the quantity $C$ for these massive particles and observed
that, contrary to might be naively expected, in thought and real
 experiments such as the one reported in \cite{Horn2},
$C_{\rm fullerenes} \sim {\cal O}(1)$.  However, for neutral particles with
permanent dipole moment this value is much lower $C_{\rm neutrons}
\sim{\cal O} (0.01)-{\cal O} (0.1)$. Therefore, on the left side of 
 Fig.\ref{graf3} we present the time
evolution of the visibility function $\nu_C(t)$ defined as
\[ \nu_C(t)=\frac{J_0(|C|)}{\cosh(8L_0 C(t)x)}.\]

Therein, we show the time evolution of the first maximum and minimum
of the interference pattern for different values of the $C$ factor.
It is easy to note that the development of the interferences happens
in the same timescale of Fig.\ref{graf2} (for the same value of
$L_0$ and $\sigma_{x0}$) but in all cases, reach a different
asymptotic value compared to the $\nu(t)$ function. The fact that
$\nu_C(t)$ has an asymptotic limit could really be of much use in
experiments where this effect is of importance, such as fullerenes, since
once this limit is reached the observation time $t_L$ can be any
subsequent time for the visibility function remains steady.

 Another feature of this visibility function $\nu_C(t)$ worth of
studying is its dependence upon the separation of the slits $L_0$.
On the right side of Fig.\ref{graf3} we present this behavior. Clearly, the behavior
exhibited therein is qualitatively different from that showed in
Fig.\ref{graf2} for the visibility function $\nu$ with $\Gamma_{\cal
D}(t)$.

\begin{figure}[t]
\includegraphics[width=8cm]{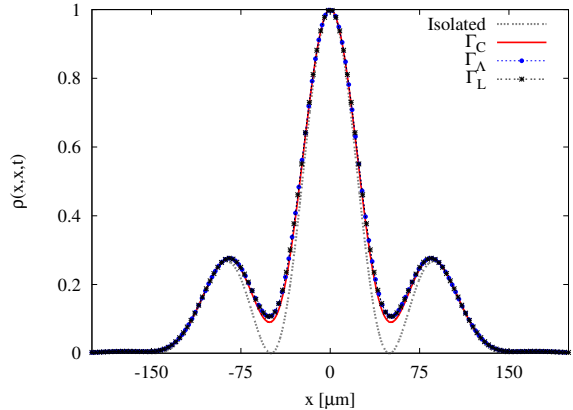}
\caption{The interference pattern ($\nu \sim 0.68$) registered on
the screen for the unitary evolution and the nonunitary evolution
considering the different model environments for a double slit
diffraction experiment with massive particles $C_{70}$. The curves
are done with the experimental data reported in \cite{Horn2}. The
values used for the plot: $\gamma_0= 2.6\times10^{-8}~s^{-1}$,
$\Lambda=2.8\times10^{15}~s^{-1}$ and $C=1$.} \label{fullerenos}
\end{figure}

Finally, in Fig.\ref{fullerenos}, the interference pattern for the
experimental data reported in  \cite{Horn2} for two-slit 
experiments with massive particles $C_{70}$ is shown. Therein, we
have considered the unitary and non unitary evolution (for the three
environmental models $\Gamma_{\cal D}(t)$, $\Gamma_{\Lambda}(t)$ and
$\Gamma_C$) of the particles. For these massive particles, we can
see that the interference pattern is always attenuated when the
system is open. What is more significant, is that the effect of
$\Gamma_C$ can be as important as the other two most widely known
model environments (in agreement with \cite{tumulka} but using a
different model for dephasing) and {\it enough} to model the real
experiment. For the values of Fig.\ref{fullerenos}, and asking $t_D
> t_L$ (and correspondingly $t_{\Lambda} > t_L$), we obtain a
constraint for the free parameters of each model:
$\gamma_0<7.14\times10^{-8}~[\rm { s^{-1}}]$ (as estimated in
\cite{Venugopalan}) and
 $\Lambda < 7.44\times 10^{15}~[\rm {m^{-2} s^{-1}}]$ (approximately
the value used in \cite{Viale}) for the experimental data at room
temperature reported in  \cite{Horn2}.

We want to emphasize that Eq.(\ref{scatering}) phenomenologically
models the decoherence effects neglecting the dissipative process.
The value we obtained for $\gamma_0$ is extremely small so a valid
question might be if it is necessary to include dissipation in the
model. We state that it positively is in order to have a
complete and formally correct description of the process. By
including the term proportional to $\gamma(t)$ in Eq.(\ref{master})
we are assuring the fulfilment of the fluctuation-dissipation
theorem, also known as Einstein formula in the high temperature
limit. It is known that the diffusive coefficient ${\cal D}= 2 M
\gamma_0 k_B T$ is proportional to $\gamma_0$. In this way, if
$\gamma_0$ happens to be zero (which means no dissipation), the
diffusive term would also be zero. The correctness of the
formulation can also be checked in the fact that even though
$\gamma_0$ is extremely small, $M \gamma_0 k_b T$ can be very large.
In such a case, the decoherence effects would be very important
whereas the dissipative interaction between the particles and the
environment can be ignored.  In other words, small dissipation
implies that the particles could have a neglible damping term in the 
semiclassical Langevin equation of motion along the ${\hat x}$ direction, but 
the existance of noise ensures that decoherence shall be effective.

It is important to stress that all these  environmental models
consider one and only one ``decoherent" agent influencing the
interference experiment. However, all these effects can be together
considered to be present in a two-slit experiment. In
such a case, the attenuation factor $\Gamma$ would be $\Gamma \approx
\Gamma_{\Lambda} + \Gamma_{{\cal D}}+ \Gamma_C$.
 The
effect of the environment would be equal to the sum of the three
factors (as seen in Fig.\ref{fullerenos}). Therefore, it is enough
to consider the biggest one (unless they are all the same order of
magnitude). All in all, it is necessary to remark that decoherence
effects do play a crucial role in the fringe visibility reduction. On the one hand, authors in
\cite{tumulka} said that the incoherence of the source was to blame
for the fringe visibility reduction. On the other hand, authors in
\cite{Sanz} developed a phenomenological theoretical model where
decoherence was {\it a priori} introduced by assuming an exponential
damping of the interferences. So far, we have shown that the scattering 
of the massive particles with
the air molecules and dephasing (for example introduced by the
random emission time) of the experimental setup are all responsible
for the fringe visibility reduction and approximately of the same order
of magnitude.

\section{Application: Experimental Data for Neutrons}
\label{aplicacion}

In this Section, we shall use the existing experimental data
\cite{Nairz,Zeilinger} to reproduce the observed patterns for
neutral cold atoms.

If we consider that $t_L=L/v=M \lambda_{\rm dB} L/(2 \pi \hbar)$ and $t_L
>> M \sigma_{x0} L_0/\hbar$, then the position distribution on the
screen at this time $t_L$ can be well approximated by: \beq
P(x,t_L)=\frac{8 \pi \sigma_{x0}^2 N^2}{\lambda_{\rm dB} L} \exp
\bigg\{-\bigg(\frac{2\sqrt{2} \pi \sigma_{x0} x}{\lambda_{\rm dB} L}
\bigg)^2\bigg\} 
\times \bigg[1 + \Gamma(t_L) \cos \bigg(\frac{2 \pi L_0
x}{\lambda_{\rm dB} L} \bigg) \bigg], \label{int}\eeq
where $\Gamma(t_L)$ depends on the model environment we want to use to
describe the conditions in which the two-slit experiment
is being done evaluated in the observation time. 
In this way, Eq.(\ref{int}) describes the intensity
on the screen as a function of the experimental parameters, i.e, the
mass $M$ of the  cold neutrons, the
associated wavelength $\lambda_{\rm dB}$, the distance to the screen
$L$, the distance between the slits $L_0$ (assuming  the two slits
are as similar as possible) and the initial width of the wave packet
$\sigma_{x0}$. All these values can be found, for example, in
\cite{Sanz} for cold neutrons. Note that Eq.(\ref{int}) is
equivalent to Eq.(\ref{prob}) , identifying our time dependent
theoretical parameters with the real experimental ones. Thus, we
have for a fixed observational time $t_L$ (making the same
assumptions as in the above section)
\[B(t_L)=\frac{2 \pi}{\lambda_{\rm dB} L} ~~~ {\rm and}~~~
C(t_L)=\bigg(
\frac{2 \sqrt{2} \pi \sigma_{x0}}{\lambda_{\rm dB} L} \bigg)^2.
\]

In the case we studied in the preceeding Section, 
assuming that the dynamics of the test
particle can be modeled by a quantum brownian motion, the expression
for $\Gamma(t_L)$ is $\Gamma_D(t_L)= \exp(-t_L/t_D)$ with $t_D=12
\hbar^2/(M \gamma_0 k_B T L_0^2)$ where we have reincorporated
$\hbar$. In the estimation of this time we have considered that $\Delta x^2
\sim L_0^2$, which in fact is an underestimation of
the decoherence time for lengths bigger than $L_0$.
As the experiment is done at room temperature, the only
free parameter is the value of $\gamma_0$. On the other hand, if the
model environment were assumed to be the one of the scattering with
air molecules, where the effects of the environment are included in
the collision term $\Lambda$, then the expression for $\Gamma(t_L)$
would be $\Gamma_{\Lambda}(t_L)= \exp(-t_L/t_{\Lambda})$ with
$t_{\Lambda}= 3/(\Lambda L_0^2)$.

\begin{figure}[t]
\includegraphics[width=8cm]{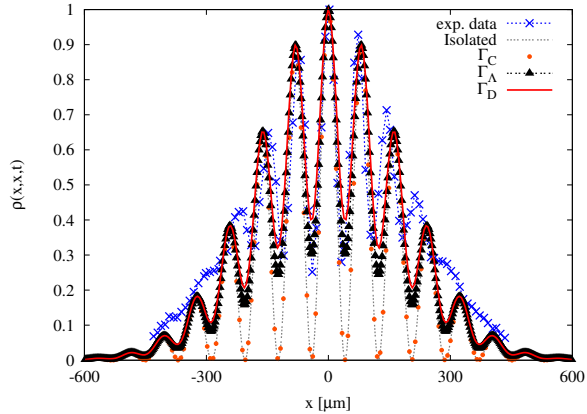}
\caption{The interference pattern ($\nu \sim 0.57$) registered on
the screen for the unitary evolution and the nonunitary evolution
considering the different model environments. Experimental results
obtained by Zeilinger et. al \cite{Zeilinger2} for the double slit
diffraction of cold neutrons are plotted with blue star dots. The
values used for the plot: $\gamma_0= 5.0\times10^{-12}~s^{-1}$ ,
$\Lambda=5.5\times10^{11}~s^{-1}$.} \label{neutrons}
\end{figure}

The other possible environmental model mentioned in Sec.\ref{intro}
was to considered the interaction with the charged or neutral (with
permanent dipole moment) particles with the electromagnetic field.
This is an interaction that is always present, can never be turned
off, although sometimes it is possible to neglect it. As we
previously studied in \cite{Nos}, in the case of neutral particles
with permanent dipole moment, this effect is not so important.

In Fig.\ref{neutrons} we have plotted, the interference pattern for
the experimental data reported for experiments
with cold neutrons
 for the isolated  and open system. In this last case,
the nonunitary evolution has been considered for the different
environment models mentioned above.  We can clearly see, that for
this case, the nonunitary evolution when considering the interaction
of the cold neutrons with a time-varying electromagnetic field
(orange dots) is exactly superposed with the unitary one (dotted
line). That means that the incoherence effects can be completely
neglected. However, the other two model environments, whose effects
are considered in $\Gamma_{\Lambda}(t)$ (black triangle dotted line)
and $\Gamma_D(t)$ (red solid line), fit correctly the experimental
data obtained by Zeilinger {\it et al.} (blue star dotted line) in
\cite{Zeilinger2}. The fact that the interference pattern is
observed implies that the decoherence time is $t_D$ (and
$t_{\Lambda}$) is larger than the observation time $t_L$. That sets
us a constraint to the expected values for $\gamma_0$ (and
$\Lambda$), the free parameter in each model. By asking $t_D>t_L$,
we have $\gamma_0 < 8\times10^{-9}~[s^{-1}]$. In the case of
modeling the environment by a collision term $\Lambda$, if
$t_{\Lambda}<t_L$ is asked, then $\Lambda < 1.28\times10^{14}~[\rm
{m^{-2} s^{-1}}]$.

\section{Final Remarks}
\label{final}

The effect of the environment on the interference pattern of a
two-slit interference experiment with massive particles has been
studied phenomenologically in the literature.

However, here we have presented a fully quantum mechanical treatment
using a microscopic model of environment and also a concrete example
to include dephasing effects. Therefore, we have studied the
effects of decoherence on the interference pattern of thought
experiments and presented an analysis of matter wave
interferometry in the presence of a dynamic quantum environment
such as the quantum brownian motion model. We have shown the
interference patterns and visibility function $\nu(t)$ for thought
diffracted free particles and analyzed their dependence upon
different parameters of the model in the high temperature limit
(assumption valid for massive particles interfering at room
temperature). As was expected, the visibility decreases as the value
of the diffusion coefficient increases and, in particular, we showed
this effect  for different values of the coupling constant
$\gamma_0$ to the environment. What is more important, we have seen
that the visibility fringe is considerably reduced when considering
an open quantum system, although the structure of the interference
pattern remains unchanged.

Yet more important, we defined the visibility function $\nu_C$ for a
model environment previously developed which describes dephasing 
effects originated in the experimental difficulty of
producing the same initial/final state for all particles (i.e the
existence of a random variable such as the particle's emission
time). We showed that it is qualitatively different than the one
commonly found in the literature and very important in the case of
experiments with massive particles such as fullerenes.
In this case, dephasing effects are enough to model the
attenuation of the interference pattern observed in the real
experiment, whereas in the case of cold neutrons  these
effects are not of such importance. Therefore, in the latter case we
must consider the decoherence effects by using the corresponding
formulation. This result might have been expected since the
interaction of more massive particles with the external classical
field is more important than for those with a smaller mass where
other kind of interactions seem to prevail.

Finally, the effect of the environment on a two-slit 
experiment can be modeled by considering different effects such as
the scattering of
the massive particles with the air molecules, the randomness of the
arrival or emission times and the presence of a classical time
dependent electromagnetic field. Even though there exist conceptual
differences in all the cases mentioned throughout the paper, we
showed that all these effects reduce the visibility fringe and can
be formally deduced from a microscopic model (whether the QBM for
decoherence effects studied in this paper or a fluctuating
Aharonov-Casher phase studied in our previous contribution). They
are all included in the noise induced effects introduced in the
subsystem when the latter is coupled to a quantum external
environment.

\section{Acknowledgments}
This work was supported by UBA, CONICET, Fundaci\'on Antorchas, and
ANPCyT, Argentina. Authors gratefully acknowledge A.S.Sanz for
sending the experimental data for neutrons. Paula I. Villar
gratefully acknowledges financial support of UIPAP.

\end{document}